# Stability Transfer between Two Clock Lasers Operating at Different Wavelengths for Absolute Frequency Measurement of Clock Transition in $^{87}$Sr

Atsushi Yamaguchi[1,2], Nobuyasu Shiga[3,1], Shigeo Nagano[1], Ying Li[1], Hiroshi Ishijima[1], Hidekazu Hachisu[1,2] Motohiro Kumagai[1], and Tetsuya Ido[1,2*]

[1]National Institute of Information and Communications Technology, Koganei, Tokyo 184-8795, Japan

[2]CREST, Japan Science and Technology Agency, Kawaguchi, Saitama 332-0012, Japan

[3]PRESTO, Japan Science and Technology Agency, Kawaguchi, Saitama 332-0012, Japan

We demonstrated transferring the stability of one highly stable clock laser operating at 729 nm to another less stable laser operating at 698 nm. The two different wavelengths were bridged using an optical frequency comb. The improved stability of the clock laser at 698 nm enabled us to evaluate the systematic frequency shifts of the Sr optical lattice clock with shorter averaging time. We determined the absolute frequency of the clock transition $^1S_0$ - $^3P_0$ in $^{87}$Sr to be 429 228 004 229 873.9 (1.4) Hz referenced to the SI second on the geoid via International Atomic Time (TAI).

---

* Email address: ido@nict.go.jp

Over the past decade, the performance of optical atomic clocks has been greatly improved.[1,2] The frequency stability of the clock laser (local oscillator) referenced to a high-finesse cavity plays a crucial role in optical atomic clock operation. Therefore, great attention has been paid to the design of the reference cavity to make it insensitive to the effects that cause frequency fluctuations.[3,4] When we operate multiple optical atomic clocks based on various clock transitions, we have to prepare such elaborate cavities for every transition, which is quite burdensome.

In this paper, we demonstrated a technique for greatly reducing this burden. We transferred the high stability of the ultrastable clock laser (master laser) to the less stable laser (slave laser) via an optical frequency comb. The slave laser was prestabilized to the cavity then phase-locked to the comb, while the comb was phase-locked to the master laser (see Fig. 1). As a result, the high stability of the master laser was transferred to the slave laser. Although a similar technique was previously demonstrated for different purposes,[5] to our knowledge, the applicability of this technique to spectroscopy of the clock transition and operation of atomic clocks has not yet been explored. We confirmed that the spectroscopy of the clock transition in $^{87}$Sr by the slave laser exhibited the same stability as that expected for the master laser.

In principle, prestabilization of the slave laser is not necessary. In that case, a wideband servo is required for both the master-to-comb and comb-to-slave links. However, a wideband servo for the master-to-comb link is usually difficult to implement because its bandwidth is normally limited to a few tens of kHz due to a piezoelectric tranceducer in a comb. As a result, the higher frequency noise of a comb is transferred to the slave laser. Although there is a technique for extending the servo bandwidth of a comb, it requires additional high-speed response components such as an intracavity electro-optical modulator.[6] There is an alternative method and, in this research, we chose to suppress the high-frequency noise of the slave laser by prestabilizing it to the cavity, and suppress only the low frequency noise via the master-to-comb and comb-to-slave links.

We first describe a Sr optical lattice clock at National Institute of Information and Communications Technology (NICT).[7] An optical lattice clock is operated with spin-polarized fermionic $^{87}$Sr atoms. An ensemble of roughly $10^4$ atoms is laser cooled to 3 μK and loaded to the vertically oriented one-dimensional (1D) optical lattice potentials. The beam waist and the potential depth of the optical lattice are 36 μm and 76 $E_r$, respectively. Here, $E_r$ is the recoil energy of the lattice laser. The clock transition (λ = 698 nm) is probed with a clock laser propagating along the strong confinement axis of the lattice potential to suppress the Doppler and recoil shifts. To excite the atoms homogeneously, we make the beam waist of the probe laser (160 μm) much larger than that of the lattice laser (36 μm). In addition, we align the probe laser parallel to the lattice laser by using a beam profiler, which leads to a residual angle of less than 0.2 mrad. The excitation ratio is determined by the fluorescence signal from excited and unexcited atoms. The spectrum of the clock transition is observed with a Fourier-limited linewidth of 19 Hz for a pulse duration of 50 ms.

Both the ground $^1S_0$ ($F = 9/2$) and excited $^3P_0$ ($F = 9/2$) states have magnetic sublevels. For the clock operation, we therefore optically pump atoms to one of the stretched states ($^1S_0$ $m_F = +9/2$ or $^1S_0$ $m_F = -9/2$) in a bias magnetic field and probe π-transitions from the two stretched states in turns in order to stabilize the clock laser to the resonance frequency for the zero magnetic field.[8,9] The applied magnetic field of 40 μT is parallel to the lattice polarization and to the propagation direction of the optical pumping laser. After optical pumping, the bias magnetic field is increased to 99 μT to reduce the line-pulling effect, though we always check that the neighboring components due to imperfect optical pumping are negligibly small. By probing both sides of the spectrum at its full width at half maximum, the deviation of the clock laser frequency from the atomic resonance is detected and used for clock operation. In a clock cycle time of 1.3 s, atoms are loaded into lattices and one side of one of the magnetic sublevels ($m_F = \pm 9/2$) is probed.

The stability transfer between two clock lasers is achieved using the system shown in Fig. 1. The master laser operating at 729 nm is the most stable clock laser at NICT, which was developed for the Ca$^+$ clock.[10] The short term stability of the master laser was characterized by the remote comparison to a clock laser in University of Tokyo (UT).[11] The result is shown in Fig. 2 as blue triangles, where the instability of the fiber link is subtracted and the stability is mainly limited by the master laser at NICT. A clock laser for the Sr optical lattice clock, described as a slave laser in Fig. 1, is a diode laser prestabilized to the optical cavity made of ultra low expansion glass. Although the finesse is 200000, the cavity is not a vibration insensitive one but a simple 10-cm cylinder. To transfer the stability of the master laser to the slave laser, the Ti:Sapphire optical frequency comb is first phase locked to the master laser.[12] The slave laser is then phase locked to the optical frequency comb (servo bandwidth of about 1 kHz). The frequency fluctuation is corrected by tuning the voltage applied to a voltage-controlled oscillator for the acousto-optical modulator (AOM) between the slave laser and frequency comb. In both phase locking stages, the digital phase frequency discriminators (PFDs) are used as mixers to extend the locking range of the phase lock.

We evaluated the improved stability of the slave laser by interleaving two independent clock stabilization programs with the same conditions. The in-loop stability calculated from the frequency difference between two data sets in the interleaved measurement is not an actual frequency stability of the clock. However, this technique is useful to evaluate the short-term stability determined by the probe laser frequency noise. The results are shown in Fig. 2. The red squares and black circles denote the stability observed with and without transferring the stability of the master laser, respectively. The corresponding dashed lines project the calculated stabilities of the slave laser: $1.9 \times 10^{-15}$ and $4.4 \times 10^{-15}$, respectively. Thus, an improvement by a factor of more than two is achieved. Here we assume that the in-loop stability is determined by Dick effect and that the noise of the slave laser can be modeled by flicker frequency noise. The red dashed line is consistent with the blue triangles in a short averaging time. This means that the technique demonstrated here faithfully transfers the short-term stability of the master laser to the slave laser without degradation.

With the improved short-term stability, we evaluated systematic shifts by using the interleaved measurement. In this measurement, the parameter to be investigated (e.g. intensity of the lattice laser) was switched between two values, synchronized with the clock cycle. The frequency shift is determined from the frequency difference between the two clock operations. Thanks to the improved stability of the clock laser, the statistical fractional uncertainty reached $6 \times 10^{-16}$ at an averaging time of only 300 s. If we do not use the stability transfer technique, the result in Fig. 2 indicates that it must take more than 1500 s to achieve the same statistical uncertainty. Using these techniques, we first evaluated the AC stark shift. From the interleaved measurements for three different lattice intensities $I_0$, 0.76 $I_0$, and 0.53 $I_0$, where $I_0$ is the lattice intensity for the normal clock operation, the AC Stark shift caused by a lattice laser was experimentally determined to be -2(2)×$10^{-16}$. We also measured the collisional shift with the same configuration. Due to the Pauli exclusion principle, the ensemble of ultracold spin-polarized fermions is beneficial for suppressing the collisional shift. From the interleaved measurements for three different atomic densities $\rho_0$, 0.8 $\rho_0$, and 0.4 $\rho_0$, where $\rho_0$ is the atomic density for normal clock operation, we experimentally determined the collisional shift to be 0.9(3.0)×$10^{-16}$.

Other systematic frequency shifts were evaluated as follows. The blackbody radiation (BBR) shift $\Delta\nu$ was estimated using $\Delta\nu$ = -2.354(32) $(T/300)^4$ Hz where $T$ is the temperature in Kelvin.[13] The room where the Sr lattice clock is located is always precisely temperature-controlled within an accuracy of 1 K. The temperature of the aluminum vacuum chamber was also measured using three calibrated thermistors. We did not observe a temperature difference of more than 1 K over the chamber. The radiation from the atomic oven was blocked by a mechanical shutter in a vacuum and a narrow metal tube for differential pumping. We estimated the overall BBR shift as -53(2)×$10^{-16}$ for a temperature uncertainty of 3 K. Using the results from other measurements, the second-order Zeeman shift caused by a bias magnetic field of 99 μT was estimated to be -5(2)×$10^{-16}$.[14] To estimate the gravitational red shift, we surveyed the height $h$ of the Sr lattice clock from the geoid surface. This height $h$ can be estimated from the equation $h = h_{ellip} - h_{geo}$, where $h_{ellip}$ and $h_{geo}$ are the heights of the Sr lattice clocks and of the geoid surface, respectively, above the earth ellipsoid based on the GRS80 reference frame. The $h_{ellip}$ is measured using Global Positioning System (GPS) survey data and $h_{geo}$ is determined using the Japanese geoid model GSIGEO2000. The $h$ for the Sr lattice clock was determined to be 76.33 m. While the uncertainties in the GSIGEO2000 and GPS survey data are both less than ten centimeters, we conservatively estimated the uncertainty as 1 m because of a few tens of centimeters discrepancy between GSIGEO2000 and the previous model as well as possible uncertainties due to tidal displacement and earthquakes.[15] Table I summarizes the frequency corrections and uncertainties.

We measured the absolute frequency of the $^{87}$Sr optical lattice clock referenced to the SI second on the geoid (TT) on April 13th, 2011. The frequency uncertainties between Sr and TT are also shown in Table I. The clock laser frequency stabilized to the Sr clock transition was measured using an optical frequency comb referenced to a hydrogen maser. The total measurement time of 26600 s leads to frequency measurement uncertainty of $17 \times 10^{-16}$. The frequency difference between the hydrogen maser and Coordinated Universal Time determined by NICT (UTC(NICT)) was monitored every second with a dual mixer time difference system.[16] The UTC(NICT) is compared with TAI via a satellite link and the time difference between them is reported in Circular T with five-day intervals. On the other hand, the Sr optical lattice clock was operated for only 26600 s in this measurement. This measurement time difference introduces an additional uncertainty of frequency transfer. We conservatively estimated the uncertainty due to the measurement time gap to be $26 \times 10^{-16}$ based on the long-term behavior of an intermediate reference, UTC(NICT). To obtain the frequency offset between UTC(NICT) and TAI, and also that between TAI and TT, we referenced the Circular T No. 280 which covered our measurement day, April 13th.[17] Finally, we determined the absolute frequency of the clock transition in $^{87}$Sr to be 429 228 004 229 873.9 (1.4) Hz. Our result agrees with values previously measured at other institutes within the uncertainty,[18-21] which indicates that our slave laser does not induce any frequency shift beyond the uncertainty.

In summary, we demonstrated a technique for transferring the stability between two clock lasers. The high stability of the master laser is faithfully transferred to the other laser. Using the improved clock laser, we could evaluate the systematic frequency shifts with shorter averaging time and determine the absolute frequency of the clock transition in $^{87}$Sr. The technique demonstrated here has a variety of applications in the field of optical frequency standards. If we use the master laser stabilized to the atomic transition, the frequency drift of the slave laser is significantly suppressed. Such an extremely stable clock laser would be advantageous for stable evaluation of certain kinds of systematic frequency shifts which requires long spectroscopy time. An optical cavity made from pure silicon which is cooled down to 120 K is currently under development to address the thermal noise[22]. Since silicon is transparent for only the infrared wavelength, stability transfer technique demonstrated here is essential to exploit its high stability for the optical clocks mostly operated in the visible range.

Acknowledgments

We thank T. Takano, M. Takamoto, and H. Katori for providing an opportunity to characterize our clock laser by the remote comparison. We are also grateful to M. Hosokawa, Y. Koyama, Y. Hanado, M. Fujieda, H. Ito, A. Nogami and K. Kido for their comments and experimental support. This research was supported in part by the JSPS through its FIRST program.

Fig. 1. Schematic diagram of system for transferring stability of master laser to slave laser (Sr clock laser). Beatnote between master laser and optical frequency comb is mixed with local oscillator (LO) by using phase/frequency discriminator (PFD) then frequency comb is phase-locked to master laser. Prestabilized slave laser is phase locked to frequency comb by applying feedback signal to AOM through voltage-controlled oscillator (VCO). Stabilized slave laser is used for spectroscopy of $^{87}$Sr.

Fig. 2. Stability of clock laser with (red square) and without (black circles) transfer of stability of master laser. Solid lines for red squares and black circles show stability of $1.0\times10^{-14}/\sqrt{\tau}$ and $2.3\times10^{-14}/\sqrt{\tau}$, respectively, where $\tau$ is averaging time. Corresponding dashed lines project calculated stabilities of slave laser. Result of remote comparison between master laser and clock laser at UT is shown as blue triangles.[11]

**Table I.** Error budget for absolute frequency measurement of clock transition in $^{87}$Sr.

| Contributor | Correction ($10^{-16}$) | Uncertainty ($10^{-16}$) |
|---|---|---|
| AC Stark (lattice) | 2 | 2 |
| AC Stark (probe) | 0.2 | 0.2 |
| Collision | -0.9 | 3 |
| Blackbody | 53 | 2 |
| 2nd Zeeman | 5 | 2 |
| Gravitational | -83 | 1 |
| Servo error | 0.0 | 0.5 |
| Sr Total | -24 | 5 |
| Sr – Maser | - | 17 |
| Maser – UTC(NICT) | - | 26* |
| UTC(NICT) - TAI | - | 9.8† |
| TAI – TT | - | 4† |
| Total | - | 33 |

*Due to short measurement time. See text. †From Circular T No. 280.

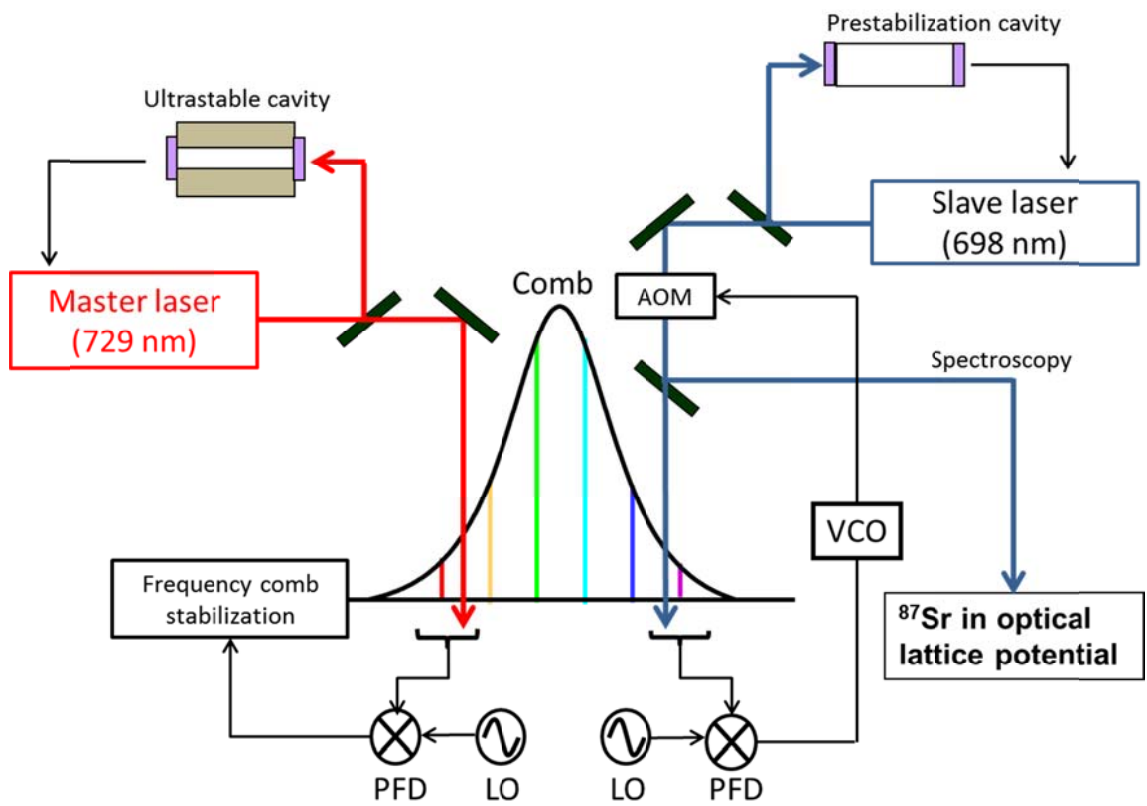

**Fig. 1.**

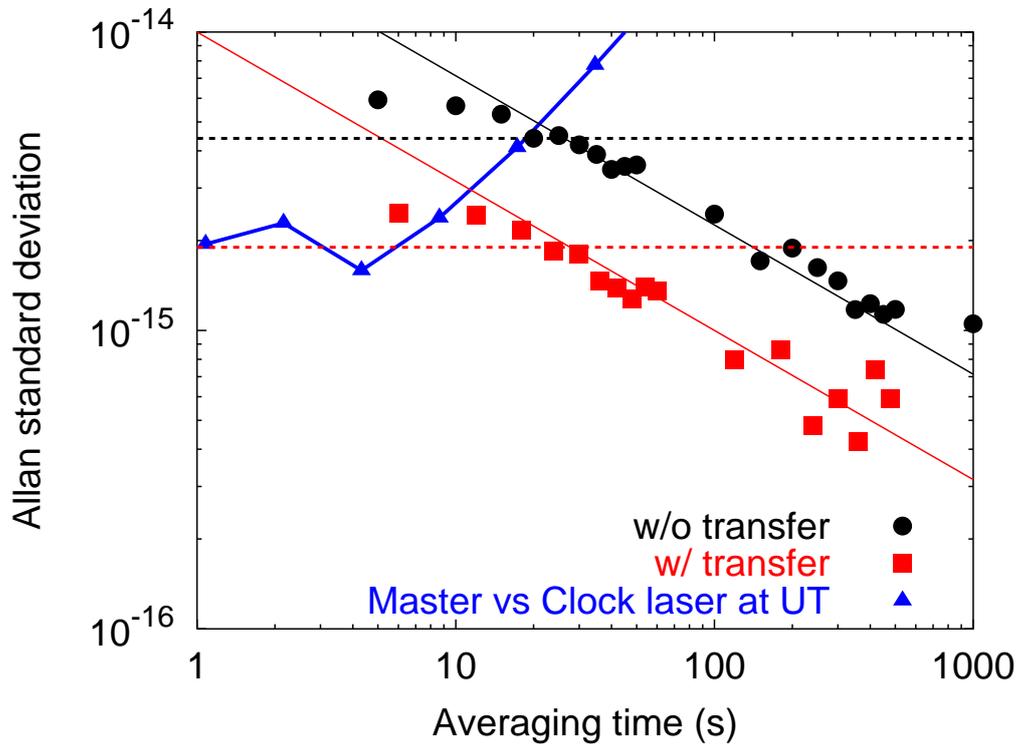

**Fig. 2.**